\documentstyle[epsf,aps,version2]{revtex}

\begin{document}

\begin{title}
\centerline{Approximate tight-binding sum rule for the superconductivity related}
\centerline{change of $c$-axis kinetic energy in multilayer cuprate superconductors}

\noindent
\author{Dominik Munzar$^{1)}$, Todd Holden$^{2)}$, Christian Bernhard$^{2)}$}

\begin{instit}
\noindent
$^{1)}$ Institute of Condensed Matter Physics,
Masaryk University,
Kotl{\'a}{\v r}sk{\'a} 2, CZ-61137 Brno, Czech Republic \\
$^{2)}$ Max-Planck-Institut f\"ur Festk\"orperforschung,
Heisenbergstra{\ss}e 1, D-70569 Stuttgart, Germany
\end{instit}

\end{title}

\begin{abstract}
We present an extension of the $c$-axis tight-binding sum rule
discussed by Chakravarty, Kee, and Abrahams [Phys.~Rev.~Lett.~{\bf 82}, 2366 (1999)]
that applies to multilayer high-$T_{c}$ cuprate superconductors (HTCS)
and use it to estimate---from available infrared data---the change below $T_{c}$
of the $c$-axis kinetic energy, $\langle H_{c}\rangle$, in
YBa$_{2}$Cu$_{3}$O$_{7-\delta}$ ($\delta=0.45\,,0.25\,,0.07$),
Bi$_{2}$Sr$_{2}$CaCu$_{2}$O$_{8}$, and
Bi$_{2}$Sr$_{2}$Ca$_{2}$Cu$_{3}$O$_{10}$.
In all these compounds $\langle H_{c}\rangle$ decreases below $T_{c}$
and except for Bi$_{2}$Sr$_{2}$CaCu$_{2}$O$_{8}$
the change of $\langle H_{c}\rangle$
is of the same order of magnitude as the condensation energy.
This observation supports the hypothesis that in multilayer HTCS
superconductivity is considerably amplified
by the interlayer tunnelling mechanism.
\end{abstract}

There is a growing evidence that
at least in some HTCS
the effective out-of-plane ($c$-axis) kinetic energy, $\langle H_{c}\rangle $,
decreases upon entering the superconducting (SC) state \cite{BasovSc,Basov1}.
In this manuscript we address the question
whether this decrease represents a significant (and perhaps the dominant)
contribution to the condensation energy
or whether it is merely a small byproduct of an in-plane pairing mechanism.
It is obvious that a quantitative estimate of the change of $\langle H_{c}\rangle $
between the normal and the SC state,
${\Delta \langle H_{c}\rangle =\langle H_{c}\rangle
(T\approx T_{c})-\langle H_{c}\rangle (T<<T_{c})}$,
is required in order to answer this interesting question.

Two different approaches have been recently employed to obtain
the value of $\Delta \langle H_{c}\rangle $:
an approximate one \cite{Anderson1998}
where $\Delta \langle H_{c}\rangle $ is identified
with the Josephson coupling energy (``JCE'', $E_{J}$)
of the internal Josephson junctions
and a rigorous one
based on the so called $c$-axis tight-binding sum rule \cite{Chakravarty1}.
Within the JCE approach, the value of $\Delta \langle
H_{c}\rangle $ per unit cell is given by the formula
$$
\Delta \langle H_{c}\rangle =E_{J}={\frac{\hbar ^{2}\varepsilon _{0}a^{2}}
{4e^{2}d}}\omega _{pl}^{2}
\eqno(1)
$$
which contains only one nontrivial parameter,
the plasma frequency of the internal Josephson plasmon $\omega _{pl}$.
The values of the in-plane lattice constant $a$
and the distance between the neighboring copper-oxygen planes $d$
are well known.
The sum-rule (SR) approach instead relates
$\Delta \langle H_{c}\rangle$
to the increase below $T_{c}$
of the low-frequency optical spectral weight (SW)
$$
\alpha (T,\omega )=\int_{0}^{\omega }\sigma _{1c}(T,\omega ^{\prime })\,
{\rm d}\omega ^{\prime }\,,
\eqno(2)
$$
$$
\Delta \langle H_{c}\rangle ={\frac{2\hbar ^{2}a^{2}}{\pi e^{2}d}}
[\alpha(T<<T_{c},\Omega _{c})-\alpha (T\approx T_{c},\Omega _{c})]
={\frac{2\hbar^{2}a^{2}}{\pi e^{2}d}}\Delta \alpha \,.
\eqno(3)
$$
Here $\Omega _{c}$ is a cutoff frequency
and $\sigma _{1c}$ is the real part of the $c$-axis conductivity $\sigma_{c}$.
What is the connection between the two approaches?
The JCE approach yields
only the contribution of the condensate
(related to the spectral weight $\rho _{s}$
of the $\delta $ peak at $\omega =0$ in $\sigma _{1c}(\omega )$)
but neglects the contribution due to the single particle tunnelling
that is related to the change $\Delta N$ of
the finite frequency part $N(T,\Omega _{c})$ of $\alpha (T,\Omega _{c})$.
The SR approach, on the other hand,
considers both contributions to $\Delta \langle H_{c}\rangle $.
Only in the limit
of vanishingly small changes at $T_{c}$
of the regular part of $\sigma _{c}$
the two approaches can be expected to yield the same result
(except for a factor of 4 as pointed out in Ref.~\cite{Chakravarty1}).
Especially the underdoped cuprate HTCS
are not very far from this limit
where the JCE approach can be expected to yield
a reasonable---better than order of magnitude---estimate of $\Delta H_{c}$.

In a previous paper \cite{ILT1}
we and our coworkers have reported the values of $E_{J}$ of two compounds
that have two copper-oxygen planes per unit cell (bilayer compounds):
YBa$_{2}$Cu$_{3}$O$_{7-\delta }$ (Y-123)
and Bi$_{2}$Sr$_{2}$CaCu$_{2}$O$_{8}$ (Bi-2212).
Note that in these compounds $E_{J}$ is determined
mainly by the frequency $\omega_{bl}$ of the intra-bilayer Josephson plasmon
\cite{Anderson1995,VdMarel1,Phonan,Gruninger,Zelezny,VdMarel2} and
by the distance between the closely spaced copper-oxygen planes.
We have shown that there is
a reasonably good agreement between the values of $E_{J}$
and the values of the condensation energy $U_{0}$
obtained from the specific heat data.
Assuming $\Delta \langle H_{c}\rangle \approx E_{J}$
we arrive at the conclusion
that the condensation energy (and the high value of $T_{c}$)
in the two compounds
can be accounted for by the change at $T_{c}$ of $\langle H_{c}\rangle $,
i.e., by the interlayer tunneling theory \cite{Anderson1997}.
Nevertheless, it can be objected
that the actual value of $\Delta \langle H_{c}\rangle $
may be much smaller than the one of $E_{J}$
and therefore should be determined by using
the complementary and more rigorous SR approach.
Unfortunately the approach in its simplest form as presented above
does not apply to multilayer compounds \cite{ILT1}.
Its derivation is based on the assumption
that the distribution of the total internal electric field
in the superconductor is homogeneous,
a condition that is not met for multilayer systems.
In the following we present an extension of the SR approach
that allows one to obtain a reasonable estimate
of $\Delta \langle H_{c}\rangle $ in multilayer cuprate HTCS.
We provide a short derivation of the key formulas
and discuss the applications to Y-123, Bi-2212, and
Bi$_{2}$Sr$_{2}$Ca$_{2}$Cu$_{3}$O$_{10}$ (Bi-2223).

For the sake of simplicity we focus on the bilayer case
and we use the notation introduced in Ref.~\cite{ILT1},
in particular in its appendix B.
The effective $c$-axis kinetic energy ($H_{c}$) of a bilayer compound
consists of the intra-bilayer term ($H_{bl}$)
and the inter-bilayer one ($H_{int}$),
$$
H_{c}=H_{bl}+H_{int}\,\,\,{\rm and}\,\,\,\Delta \langle H_{c}\rangle =\Delta
\langle H_{bl}\rangle +\Delta \langle H_{int}\rangle \,.
\eqno(4)
$$
The current densities in the intra-bilayer region
and in the inter-bilayer region
denoted as $j_{bl}$ and $j_{int}$, respectively,
are related to the two electric fields $E_{bl}$ and $E_{int}$
as follows (Eq.~(B7) of Ref.~\cite{ILT1}):
$j_{\alpha }(\omega )=\sigma _{\alpha \,\beta }(\omega )E_{\beta}(\omega )$,
$\alpha ,\,\beta \,\in \{bl,\,int\}$.
It is reasonable to assume that the conductivity matrix $\sigma _{\alpha \,\beta }$
is diagonal and to introduce the abbreviations $\sigma _{bl}=\sigma _{bl\,bl}$
(``intra-bilayer conductivity'') and $\sigma _{int}=\sigma _{int\,int}$
(``inter-bilayer conductivity'') so that $j_{bl}=\sigma _{bl}E_{bl}$ and $%
j_{int}=\sigma _{int}E_{int}$. The total conductivity is then given as
follows:
$$
\sigma (\omega )=(d_{bl}+d_{int})/\left[ {\frac{d_{bl}}{\sigma _{bl}(\omega )%
}}+{\frac{d_{int}}{\sigma _{int}(\omega )}}\right] \,,
\eqno(5)
$$
which is the formula used at the phenomenological level
introduced in Ref.~\cite{VdMarel1}.
We aim at finding a formula connecting $\Delta \langle H_{c}\rangle $
and $\Delta \alpha $ as defined in Eq.~(3).
Let us first express the two components
$\Delta \langle H_{bl}\rangle $ and $\Delta\langle H_{int}\rangle $
of $\Delta \langle H_{c}\rangle $
in terms of the two conductivities.
It follows from Eqs.~of Appendix B of Ref.~\cite{ILT1} that
$$
\Delta \langle H_{bl}\rangle ={\frac{2\hbar ^{2}a^{2}}{\pi e^{2}d_{bl}}}%
\Delta \alpha _{bl}\eqno(6{\rm a})
$$%
and
$$
\Delta \langle H_{int}\rangle ={\frac{2\hbar ^{2}a^{2}}{\pi e^{2}d_{int}}}%
\Delta \alpha _{int}\,,\eqno(6{\rm b})
$$%
where $\alpha _{bl/int}$ and $\Delta \alpha _{bl/int}$ are related
to $\sigma _{bl/int}(\omega )$ in the same way
as $\alpha $ and $\Delta \alpha $ are to $\sigma _{c}(\omega )$.
A schematic representation of
$\Delta \alpha_{bl}$ and $\Delta \alpha _{int}$ is shown in Fig.~1.
The quantity $\Delta\alpha $ can also be expressed
in terms of $\Delta \alpha _{bl}$ and $\Delta\alpha _{int}$.
After some manipulations
(similar to those in chapter 5.7 of Ref.~\cite{Mahan})
using the analytic and the asymptotic properties of
$\sigma (\omega )$, $\sigma _{bl}(\omega )$ and $\sigma _{int}(\omega )$
we obtain
\[
\alpha (T,\omega \rightarrow \infty )={\frac{d_{bl}}{d_{bl}+d_{int}}}\alpha
_{bl}(T,\omega \rightarrow \infty )+{\frac{d_{int}}{d_{bl}+d_{int}}}\alpha
_{int}(T,\omega \rightarrow \infty )\,.
\]%
Provided that the temperature dependence of the three conductivities
above $\Omega _{c}$ is negligible we can also write
$$
\Delta \alpha ={\frac{d_{bl}}{d_{bl}+d_{int}}}\Delta \alpha _{bl}+
{\frac{d_{int}}{d_{bl}+d_{int}}}\Delta \alpha _{int}\,.\eqno(7)
$$
In order to obtain an estimate of $\Delta \langle H_{c}\rangle $
we have to make an additional assumption because the Eqs.~(4), (6a), (6b) and (7)
still contain five unknowns: $\Delta \langle H_{c}\rangle $,
$\Delta \langle H_{bl}\rangle $, $\Delta \langle H_{int}\rangle $,
$\Delta \alpha _{bl}$ and $\Delta \alpha _{int}$
($\Delta \alpha $ is assumed to be known from the infrared data).
We suggest to use the following assumption:
$$
{\frac{\Delta \alpha _{bl}}{\Delta \alpha _{int}}}={\frac{\omega _{bl}^{2}}{%
\omega _{int}^{2}}}\,,\eqno(8)
$$%
where $\omega _{bl}$ and $\omega _{int}$ are the superfluid plasma
frequencies of the intra-bilayer region and the inter-bilayer region,
respectively. In other words, we suggest that $\Delta \alpha _{bl/int}$
is proportional to the contribution $\rho _{s\,bl/int}$ of the
condensate (see Fig.~1 for a definition of $\rho _{s\,bl/int}$).
Using Eqs.~(4), (6a), (6b), (7) and (8) we finally arrive at the formula
expressing $\Delta \langle H_{c}\rangle $ in terms of $\Delta \alpha $:
$$
\Delta \langle H_{c}\rangle =k\,\,{\frac{d_{bl}+d_{int}}{d_{bl}d_{int}}}\,\,{%
\frac{\omega _{bl}^{2}d_{int}+\omega _{int}^{2}d_{bl}}{\omega
_{bl}^{2}d_{bl}+\omega _{int}^{2}d_{int}}}\,\,\Delta \alpha \,,\eqno(9)
$$%
where $k=2\hbar ^{2}a^{2}/(\pi \,e^{2})$. There is one special case, where
there is no need to use any additional assumption like Eq.~(8),
the case of negligible inter-bilayer conductivity.
Then $\Delta \alpha _{int}$ can be neglected and we obtain
$$
\Delta \langle H_{c}\rangle =k\,\,{\frac{d_{bl}+d_{int}}{d_{bl}^{2}}}
\,\,\Delta \alpha \,\,\,
{\rm or\,equivalently}
$$
$$
\Delta H_{c}[{\rm meV}]={%
\frac{4.7\cdot 10^{-5}}{d_{bl}[\text{\AA}]}}
{\frac{d_{bl}+d_{int}}{d_{bl}}}\Delta \alpha \lbrack \Omega ^{-1}
{\rm cm^{-2}}]\,.\eqno(10)
$$
The following points deserve some comments.\newline
(i) Applicability of Eq.~(10): Equation (10) can certainly be used for
systems like Bi-2212, where the inter-bilayer conductivity is known to be
very small. However, it can still be used---with a precision of about $20\%$%
---for the less anisotropic Y-123
because even there the ratio $\omega_{int}^{2}/\omega_{bl}^{2}$ as
obtained from the infrared data is fairly small, especially in the
underdoped samples.\newline (ii) Comparison with the sum-rule for
single layer materials: Note the difference between the right hand
sides of Eq.~(3) and Eq.~(10). The right hand side of Eq.~(10) is
larger by a factor of $(d_{bl}+d_{int})^{2}/d_{bl}^{2}$ than the
one of Eq.~(3) (with $d=d_{bl}+d_{int}$). It means that the
sum-rule in Eq.~(3) underestimates $\Delta \langle H_{c}\rangle $
by a factor of $d_{bl}^{2}/(d_{bl}+d_{int})^{2}$ ($1/20$ for
Bi-2212). {\it For a strongly anisotropic bilayer compound even a
tiny change of SW may correspond to a significant change of
$\langle H_{c}\rangle$}.
\newline
(iii) Interband polarizability:
In the derivation of Eq.~(10) we have tacitly assumed
that the interband polarizabilities of the two regions,
intra-bilayer and inter-bilayer, are the same, i.e.,
that the two conductivities, $\sigma _{bl}$ and $\sigma _{int}$,
contain the same term $-i\omega \varepsilon _{\infty }$.
For $\varepsilon _{\infty \,bl}\not=\varepsilon _{\infty \,int}$
we would obtain
$$
\Delta \langle H_{c}\rangle =k\,{\frac{\Delta \alpha _{bl}}{d_{bl}}}=k\,\,{%
\frac{(d_{bl}\varepsilon _{\infty \,int}+d_{int}\varepsilon _{\infty
\,bl})^{2}}{d_{bl}^{2}(d_{bl}+d_{int})\varepsilon _{\infty \,int}^{2}}}%
\,\,\Delta \alpha \,.\eqno(11)
$$%
(iv) SR for multilayer cuprates with $n>2$:
The approach presented above can
easily be generalized and we obtain
$$
\Delta \langle H_{c}\rangle =k\,\,{\frac{(n-1)d_{bl}+d_{int}}{d_{bl}^{2}%
}}\,\,\Delta \alpha \,.\eqno(12)
$$%
The distance between the closely-spaced copper-oxygen planes
is denoted by $d_{bl}$,
$d_{int}$ is the distance between the multilayer blocks.
In order to obtain Eq.~(12) we have assumed that
the inter-multilayer conductivity is negligible
and that all the ``Josephson junctions''
(regions between neighboring copper-oxygen planes)
within the multilayer block
exhibit the same electronic conductivity.

In Table I we show the values of $\rho _{s}$, $\Delta N$,
$\Delta \alpha=\rho _{s}+\Delta N$,
$\Delta \langle H_{c}\rangle $, $E_{J}$,
and the condensation energy per unit cell ($U_{0}$)
for Y123, Bi-2212, and Bi-2223.
\newline
\underline{$\rho _{s}$}~~The values of $\rho _{s}$ for Y-123 have been obtained
by using Eq.~(A3) of Ref.~\cite{ILT1}
and the values of $\omega _{bl}$ and $\omega _{int}$
presented in Refs.~\cite{Phonan,Gruninger}.
The corresponding values of the $c$-axis plasma frequency
are $250$, $520$, and $1500{\rm \,cm^{-1}}$.
The values of $\rho _{s}$ in the Bi-compounds are negligibly
small.\newline
\underline{$\Delta N$}~~$\Delta N=N_{s}(\Omega _{c})-N_{n}(\Omega _{c})$,
$N_{s}(\omega )=N(T<<T_{c},\omega )$, $N_{n}=N(T\approx T_{c},\omega )$,
${N(T,\omega )=\int_{0^{+}}^{\omega }\sigma _{1c}(T,\omega ^{\prime })\,{\rm d}
\omega ^{\prime }}$.
The values of $\Delta N$ for underdoped Y-123 have been obtained
by integrating the conductivity data presented in Refs.~\cite{Phonan,Bernhard}
supplemented by mid-infrared data ranging up to $\Omega_{c}=1500{\rm \,cm^{-1}}$
that have been more recently obtained by ellipsometric measurements.
A linear extrapolation of $\sigma_{1c}$ below $100{\rm \,cm^{-1}}$ has been used.
Figure 2(a) shows a complete set of ellipsometric data
for the $T_{c}=80{\rm \,K}$ sample
including the extrapolation.
Note that $\sigma _{1c}(T=20{\rm \,K}<<T_{c})$ is larger than
$\sigma _{1c}(T=100{\rm \,K}\approx T_{c})$ in two different regions:
in the region around $550{\rm \,cm^{-1}}$ (label $B$ in Fig.~2(a))
and in the one located above the frequency range
of the apical oxygen modes and centered at $900{\rm \,cm^{-1}}$ (label $C$).
The additional absorption band $B$
has already been attributed \cite{Phonan,Gruninger}
to the transverse plasma excitation (TPE)\cite{VdMarel1,Phonan,Gruninger,VdMarel2}.
We propose that the band $C$ also belongs to the TPE
since a splitting of the SW of the TPE into
two parts---one below and
one above the frequency range of the apical-oxygen modes---is
consistent with the Josephson superlattice model (JSM)
\cite{VdMarel1,Phonan,Gruninger,VdMarel2}.
This is demonstrated in the inset of Fig.~2(a)
which shows results of the model calculations of $\sigma _{1c}$
for two sets of parameters,
one corresponding to the SC state and one to the normal state.
Details of the calculations are given in Ref.~\cite{Parameters}.
The frequency dependence of the quantity $N_{n}(\omega )-N_{s}(\omega )$
corresponding to the data of Fig.~2(a) is shown in Fig.~2(b).
Note that our value of $|\Delta N|$ of $0.5\rho _{s}$
is considerably smaller than the one of $0.8\rho _{s}$
obtained by Basov {\it et al.} \cite{Basov1} for samples
with a similar value of $T_{c}$.
This is mainly due to the fact that the band $C$
appears above the value
of $\Omega _{c}$ of Ref.~\cite{Basov1} ($800{\rm \,cm^{-1}}$)
while below our value of $1500{\rm \,cm^{-1}}$.
The value of $\Delta N$ for optimum doped Y-123
has been estimated from the inset of Fig.~1 of Ref.~\cite{Gruninger}.
The difference with respect to the result presented in Ref.~\cite{Basov1}
is again related to the difference
in the value of the cutoff frequency $\Omega _{c}$.
The values of $\Delta N$ for Bi-2212 and Bi-2223
are taken from Refs.~\cite{ILT1} and \cite{Boris}, respectively. \newline
\underline{$\Delta \langle H_{c}\rangle $}~~The values of $\Delta \langle
H_{c}\rangle $ result from Eq.~(9) (Y-123), Eq.~(10) (Bi-2212), and Eq.~(12)
(Bi-2223). \newline
\underline{$U_{0}$}~~The values of $U_{0}$ have been obtained from the specific heat
data in Ref.~\cite{Loram1} (Y123) and in Ref.~\cite{Loram2} (Bi-2212),
the value for Bi-2223 is
an estimated upper limit, ${U_{0}({\rm Bi-2223})\approx 2U_{0}({\rm Bi-2212})}$.

In all the compounds studied
$\langle H_{c}\rangle $ decreases below $T_{c}$
and in all of them except for Bi-2212
$\Delta \langle H_{c}\rangle $
is of the same order of magnitude as $U_{0}$.
The values of $\Delta \langle H_{c}\rangle $ in Y-123
are smaller (ca by a factor of 2) than those of $E_{J}$
presented in Ref.~\cite{ILT1},
whereas according to the simplest version of the JSM,
neglecting the single particle tunnelling,
they should be larger by a factor of 4 \cite{Chakravarty1,ILT1}.
This is related to the fact
that the values of the SW change $\Delta \alpha$
are considerably smaller (ca by a factor of 8)\cite{SWofTPE}
than the estimates of the SW of the TPE
based on the JSM.
We are aware of the following reasons for this discrepancy.\newline
(i) Non zero single particle contribution, i.e.,
$\Delta N_{int}<0$ and $\Delta N_{bl}<0$
(see Fig.~1 for a definition of $\Delta N_{int}$ and $\Delta N_{bl}$).
Physically this means that parts
of $\rho _{s\,int}$ and $\rho _{s\,bl}$
originate from the low-frequency regions
of $\sigma _{int\,1}$ and $\sigma _{bl\,1}$, respectively.
As a consequence, the SR-based estimate
of $\Delta \langle H_{c}\rangle $
determined by the changes of the total FIR SWs,
$\Delta \alpha _{int}<\rho _{s\,int}$
and $\Delta \alpha _{bl}<\rho_{s\,bl}$, (see Eqs.~(6a) and (6b))
is smaller than the JCE estimate determined solely
by $\rho _{s\,int}$ and $\rho _{s\,bl}$.
This is probably the main reason for the discrepancy. \newline
(ii) Finite compressibility (FC) effects.
The latter have been shown \cite{VdMarel2}
to shift both the frequency of the (longitudinal) intra-bilayer plasmon
and the frequency of the TPE towards higher values
with respect to the ``bar values'' of the simplest version of the JSM.
At the same time the FC effects do not influence the SW of the TPE.
Since the effects have not been considered when analyzing the data
in Refs.~\cite{Phonan,Gruninger,Zelezny,ILT1},
the resulting values of $\omega _{bl}$ presented therein,
and collected in Ref.~\cite{ILT1},
may be a bit higher than the actual ones.
This would imply that also the values of $E_{J}$ in Ref.~\cite{ILT1}
are higher than the actual ones,
which would account at least for a part of the discrepancy.
We do not think, however, that this is the main reason for it.
If the influence of the FC effects was considerable,
the frequency of the TPE in Bi-2223
would be significantly lower than in Bi-2212
because the FC induced shift of the plasma frequencies
decreases with increasing distance
between the outer copper-oxygen planes of the multilayer block.
This has not been observed \cite{Zelezny,ILT1,Boris}.%
\newline
(iii) ``Pseudogap below $T_{c}$''.
The difference between the results of the two approaches (JCE and SR)
could be much smaller, if we considered
the temperature evolution of the normal state spectra below $T_{c}$,
i.e., if $\Delta \alpha $ was taken as
$\alpha (T<<T_{c},\Omega _{c})-{\rm lim}_{T\rightarrow 0}
\alpha ({\rm normal\,\,state},T,\Omega _{c})$.
It is this limit that should be actually used when estimating
$\Delta \langle H_{c}\rangle $
as emphasized also elsewhere \cite{Basov1}.
\newline
(iv) Fluctuation effects above $T_{c}$.
In strongly underdoped Y-123 the additional absorption peak
starts to form already at temperatures much higher than $T_{c}$
\cite{Homes,Bernhard}, presumably due to fluctuation effects \cite{Phonan,ILT1}.
The JCE-based estimate of $\Delta \langle H_{c}\rangle $
contains a contribution of these effects since it is
determined by the low temperature value of $\omega _{bl}$
(i.e., in a way that does not require
any assumptions concerning the onset of superconductivity).
On the other hand the effects obviously do not
contribute to the SR-based estimate determined
by the change of the spectra below $T_{c}$.

The SW of the additional absorption band corresponding to the TPE
($\Delta \alpha$ in Table 1)
is suprisingly small in Bi-2212 as compared to Y-123.
We suggest that this is largely due to the fact
that the spacing layer in Bi-2212 is more insulating than that of Y-123.
Indeed, it can be seen from Eq.~(11)
that for a given value of $\Delta \alpha_{bl}$
a decrease of the ratio $\varepsilon_{\infty\,int}/\varepsilon_{\infty\,bl}$
leads to a reduction of $\Delta \alpha$.
A similar reduction of $\Delta \alpha$ can also be caused
by the electronic background of the intra-bilayer region: this is
what has been assumed in Refs.~\cite{Zelezny,ILT1}.
These observations indicate
that the estimate of $\Delta \langle H_{c}\rangle$ in Bi-2212
based on Eq.~(10)
may be considerably lower than the actual value of this quantity.
Furthermore, it can not be excluded yet that a part of $\Delta \alpha$
appears at frequencies above the FIR range.
At present we do not know why $\Delta \alpha$
is almost by a factor of 5 larger in Bi-2223 than in Bi-2212
(the JSM yields a factor of ca 2).

In summary, we have developed an extension
of the $c$-axis tight-binding SR that applies to multilayer HTCS
and allows one to estimate---model independently---
the kinetic energy change $\Delta\langle H_{c}\rangle$
associated with the SC transition.
For multilayer HTCS with insulating (or almost insulating) spacing layers
the ratio between the SW change and $\Delta \langle H_{c}\rangle$
is determined by a geometrical factor that
is typically an order of magnitude lower
than the one of the conventional tight-binding SR.
Using published far-infrared data
that are in part complemented
by new MIR data for underdoped Y-123,
we found that
$\langle H_{c}\rangle$ decreases below $T_{c}$ in all the compounds studied
including optimally doped Y-123 and almost optimally doped Bi-2212 and Bi-2223.
The decrease seems thus not to be restricted to the underdoped regime.
In all the compounds studied except for Bi-2212
$\Delta \langle H_{c}\rangle$ as determined by the SR
is of the same order of magnitude as the condensation energy
and there are several reasons to believe
that for Bi-2212 $\Delta \langle H_{c}\rangle$
is underestimated.
To conclude, in all the multilayer HTCS studied
the decrease of $\langle H_{c}\rangle$ below $T_{c}$ does represent
a significant contribution to the condensation energy.
Its high value suggests that it may well be the dominant contribution.
The possibility, however,
that the changes of $\langle H_{c}\rangle $ below $T_{c}$
simply parallel much larger changes of the in-plane kinetic energy
(such as obtained in Ref.~\cite{Molegraaf} for Bi-2212)
cannot be excluded.

We acknowledge stimulating discussions with D.~van der Marel.
T.~H.~thanks the AvH Foundation for support.

\newpage
\begin{table}[tbp]
\caption{ Values of the superfluid spectral weight $\protect\rho_{s}$, the
quantity $\Delta N$ defined in the text, the spectral weight change $\Delta
\protect\alpha$, the kinetic energy change $\Delta \langle H_{c}\rangle$,
the Josephson coupling energy $E_{J}$ as obtained in Ref.~\protect\cite{ILT1}%
, and the condensation energy $U_{0}$. The quantities $\protect\rho_{s}$, $%
\Delta N$, and $\Delta \protect\alpha$ are given in ${\rm m}\Omega^{-1}{\rm %
cm}^{-2}$, the quantities $\Delta \langle H_{c}\rangle$, $E_{J}$, and $U_{0}$
in meV. }\vskip 0.3 in
\begin{tabular}{cccccccccc}
{compound} & \multicolumn{1}{c}{$T_{c} [{\rm K}]$} & \multicolumn{1}{c}{$%
\rho_{s}$} & \multicolumn{1}{c}{$\Delta N$} & \multicolumn{1}{c}{$-\Delta
N/\rho_{s}$} & \multicolumn{1}{c}{$\Delta \alpha$} & \multicolumn{1}{c}{$%
\Delta \langle H_{c}\rangle $} & \multicolumn{1}{c}{$E_{J}$} &
\multicolumn{1}{c}{$U_{0}$} & \multicolumn{1}{c}{$\Delta \langle H_{c}\rangle
/U_{0}$} \\
\tableline YBa$_{2}$Cu$_{3}$O$_{6.55}$ & 53 & 1.7 & 0 & 0 & 1.7 & 0.08 & 0.13
& 0.05 & 1.6 \\
\tableline YBa$_{2}$Cu$_{3}$O$_{6.75}$ & 80 & 7.2 & -3.5 & 0.5 & 3.7 & 0.16
& 0.30 & 0.16 & 1.0 \\
\tableline YBa$_{2}$Cu$_{3}$O$_{6.93}$ & 91 & 60.0 & -48.0 & 0.8 & 12.0 &
0.47 & 1.14 & 0.36 & 1.3 \\
\tableline Bi$_{2}$Sr$_{2}$CaCu$_{2}$O$_{8}$ & 91 & 0 & 0.3 & --- & 0.3 &
0.02 & 0.13 & 0.13 & 0.15 \\
\tableline Bi$_{2}$Sr$_{2}$Ca$_{2}$Cu$_{3}$O$_{10}$ & 107 & 0 & 1.4 & --- &
1.4 & 0.11 & - & 0.26 & 0.4%
\end{tabular}
\vskip 0.1 in
\end{table}

\figure{
Schematic representation of (a) the bilayer geometry and
(b) the spectra of $\sigma_{int\,1}=Re\{\sigma_{int}\}$
and $\sigma_{bl\,1}=Re\{\sigma_{bl}\}$
and the quantities that describe the related spectral weight changes
(as discussed in the text).}

\figure{
(a) Experimental spectra of the $c$-axis conductivity
of slightly underdoped ($T_{c}=80{\rm\,K}$) Y-123.
The labels $A$, $B$, and $C$ indicate
the region of a pronounced gap-like suppression of $\sigma_{1c}$,
the main part of the additional absorption band
due to the transverse plasma excitation,
and its high frequency satellite, respectively.
Inset: results of model calculations for the superconducting state
(solid line) and for the normal state (dashed line)
demonstrating the splitting of the additional absorption
band. (b) Spectra of the quantity $N_{n}-N_{s}$ defined in the text.
}

\newpage
\centerline{Fig.~1}
\epsffile{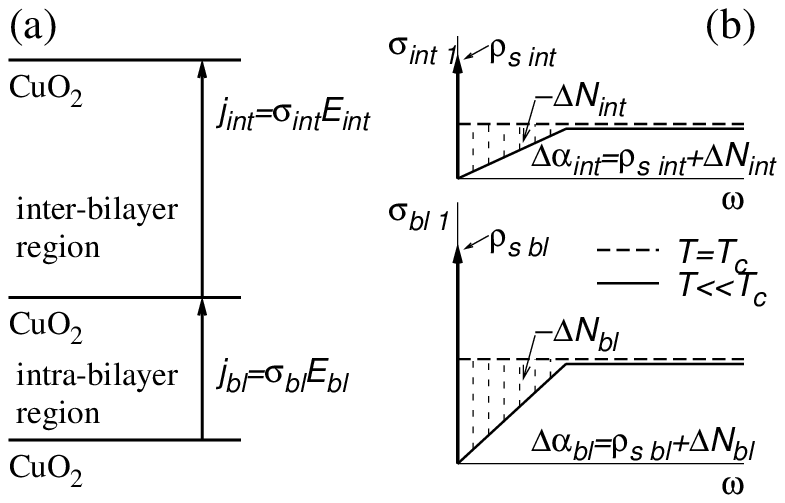}

\newpage
\centerline{Fig.~2}
\epsffile{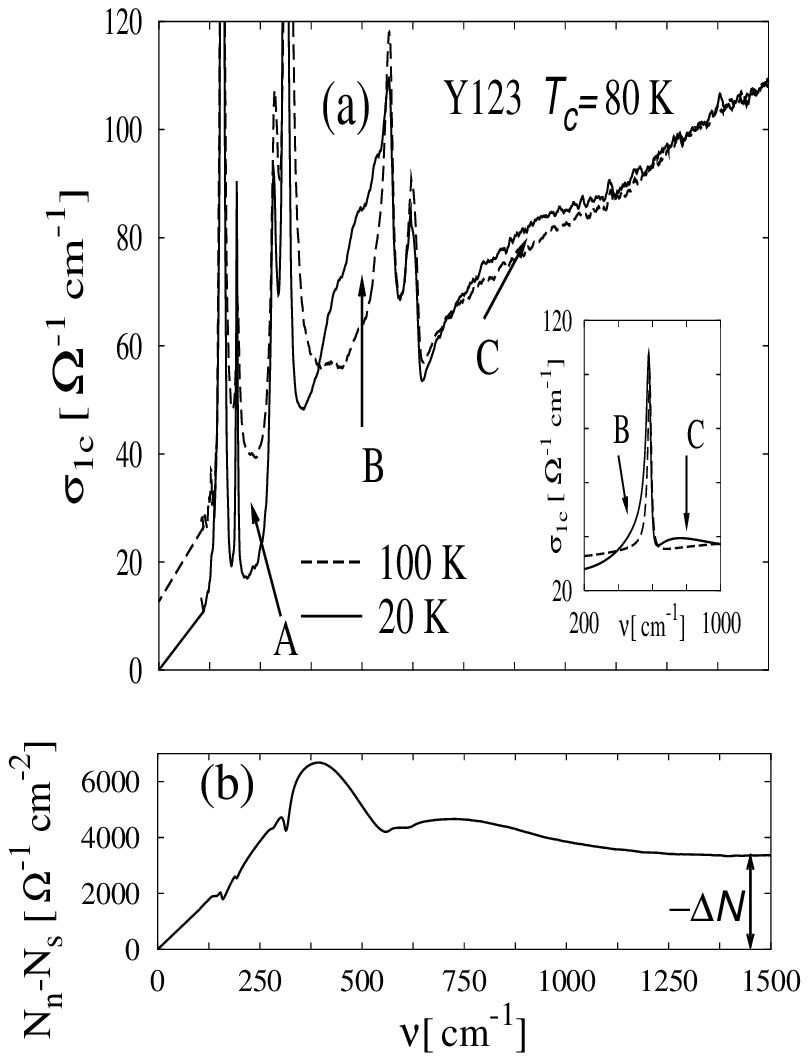}

\begin{references}
\bibitem{SWofTPE} The total SW of the additional absorption peak
at $410 {\rm\,cm^{-1}}$ in underdoped Y-123 with $T_{c}=53{\rm\,K}$,
however, is fairly close to the estimate based on the JSM model.
\end{references}
\end{document}